# Conceptual Temporal Modeling Applied to Databases

Sabah Al-Fedaghi
Computer Engineering Department, Kuwait University, Kuwait

*Abstract*—We present a different approach to developing a concept of time for specifying temporality in the conceptual modeling of software and database systems. In the database field, various proposals and products address temporal data. The difficulty with most of the current approaches to modeling temporality is that they represent and record time as just another type of data (e.g., values of a bank balance or amounts of money), instead of appreciating that time and its values are unique, in comparison to typical data attributes. Time is an engulfing phenomenon that lifts a system's entire model from staticity to dynamism and beyond. In this paper, we propose a conceptualization of temporality involving the construction of a multilevel modeling method that progresses from static representation to system compositions that form regions of dynamism. Then, a chronology of events is used to define the system's behavior. Lastly, the events are viewed as data sources with which to build a temporal model. A case-study model of a temporal banking-management system database that extends UML and the object-constraint language is re-modeled using thinging machine (TM) modeling. The resultant TM diagrammatic specification delivers a new approach to temporality that can be extended to be a holistic monitoring system for historic data and events.

## I. Introduction

In most existing relational database systems, data objects are stored such that when an attribute's value changes, the new value replaces the old value. Thus, only the latest state of an object resides in the database. However, discarding old information is inappropriate for many database applications (e.g., financial, health-care management, reservation, medical, and decision support system applications). In these cases, time values must be associated with data to indicate the time for which the data are valid. A time dimension is added to a database at either the attribute or the tuple level to maintain a data object's history. Such a database is referred to as a temporal database [1]. Many temporal extensions of the classical relational database model have been proposed, and some of them have been realized (e.g., [2], [3]).

Conceptual models are essential for describing an application's requirements, and they facilitate communication between users and designers because they do not require knowledge of the technical features of the underlying implementation platform [4]. A conceptual model provides a notation and formalism, which designers can use to construct a high-level, implementation-independent description of selected aspects of the modeled portion of reality [5].

Time is a source of mystery and sometimes is treated as a philosophical curiosity [6]. In software engineering, modeling research mostly adopts a "clock-based" mechanistic interpretation of time and ignores the complex, multifaceted, subtle, and socially embedded nature of temporality [7]. Proposals and products have been developed in the database field to address temporal data (e.g., SQL/temporal) [8]. Various time-related concepts exist (e.g., according to Halpin [8]), and three basic temporal data types may be distinguished: instant, interval, and period. Temporal operators (e.g., subtraction) can be defined for each of these types. Temporal database terminology includes many kinds of time values, including valid time, transaction time, snapshot, bitemporal, spun, and time stamp. Extra columns in relational tables often capture some of these times, and the time-as-value sometimes is distinguished from the facts (true propositions). According to Halpin [8], facts are completely devoid of any temporal aspect, and once-only facts correspond to a single event, for which an event seems to be defined as a state of affairs.

### A. Related Works

Almost all current approaches to data temporality take the same conceptualization scheme described in the previous paragraph. Surveying more such works would not serve our purposes because what we propose in this paper is not based on their paradigm. Instead, we describe the difference between an example how such approaches conceptualize temporality and our proposed conceptualization of data temporality in our thinging machine (TM) methodology.

### B. Difficulties

Over the years, the topic of temporality has been a rich research theme, such as in philosophical essays on time and temporal reasoning. Time is an important notion in many real-world applications. In recent years, research on data temporality has spread to other areas (e.g., the temporal dimensions of semantic Web applications and temporal ontologies). Yet, the field of temporality studies for such applications lacks a common terminology, infrastructure, and conceptual framework, thus reducing the adoption of temporal database technology [9]. According to Lu et al. [10], we have witnessed a major burst of temporal support in conventional database management systems; however, the existing temporal data model is inadequate, and such databases suffer from limited expressiveness [10]. Additionally, notions such as the transaction time in temporal data models are difficult to establish and update [10].

### C. Proposed Solution to the Identified Cause

In this paper, we advocate the thesis that the difficulties with current approaches to representing and recording time notion originate from them handling time data as just another value, similar to the number of cars, the length of a tree, or an amount of money. This conception reflects a lack of appreciation for time as a one-of-a-kind singularity distinct from things such as attributes. In this study, we view time as an engulfing phenomenon that lifts the entire model description from staticity to dynamism. Such movement from staticity converts static description into eventized form (TM machine), injecting activity into the whole system in a way that is analogous to transforming a mere textual narrative into a live theater performance. The text, "Alice comes across a caterpillar sitting on a mushroom", when converted to a performance in a theater, becomes Alice as a thing knotted in time comes across as an action knotted in time, a caterpillar as a thing knotted in time, sitting as an action knotted in time, on a mushroom as a thing knotted in time. The sentence "Alice comes across a caterpillar sitting on a mushroom", with its things and action, is a static model, whereas the performance is a dynamic model in which things and actions are transformed into events. Our model captures such a distinction between staticity and dynamism that is not found in most other models (e.g., UML) representing dynamism in a static form (e.g., a UML activity diagram).

The static representation shown in Fig. 1 (adapted from [11]), in which the time is the value of the temporal attribute Date, lacks the key notion that time data are generated at a higher level than that of static descriptions. This idea is adopted in this paper, whereby a static fact (*Employee*, *MoneyAmount*) is at a lower modeling level than time is (Fig. 2). Fig. 2 shows the lower level (dark region) and upper level, along with their respective data (e.g., Date).

*D. Outline of the Approach*

In this paper, we express the domain involved (i.e., bank-operations management) in terms of a new modeling methodology called TM modeling, which is a conceptual tool that abstractly represents a system at four levels: the static, dynamic, behavioral, and temporal levels. The crucial construction of this multilevel modeling involves a progression from the static representation of a system to a set of compositions that form regions of events. The event chronology defines the system behavior. Lastly, the events are viewed as data sources for a temporal system.

These modeling stages are illustrated in a simplified form using a single decomposition (Fig. 3). Fig. 4 shows the corresponding event. Fig. 5 shows the meta-event that produces salary data and its data of occurrence, which is stored as a record in the temporal salary database. Of course, the record includes other data such as employee names and ID numbers. This record is generated each time the event of updating the salary value occurs.

To provide a self-contained paper, the next section includes a brief summary of the TM model. Section 3 details a full example of TM modeling for understanding the model and its concepts, such as actions, events, and behaviors. Section 4 introduces a temporal database given by El Hayat et al. [12] that extends UML and the object-constraint language (OCL) to produce an elaborate conceptual schema of banking-system management. This temporal database is re-modeled using the TM methodology.

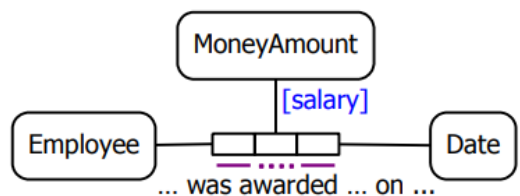

Fig. 1. Model of a Salary with History (Adapted from Balsters et al. [11]).

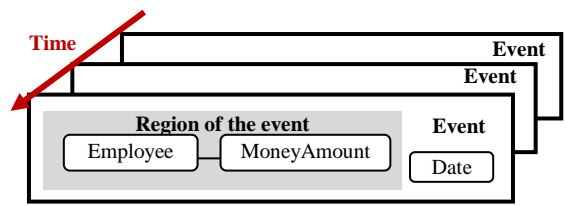

Fig. 2. The Conceptualization of Temporality Adopted in this Paper.

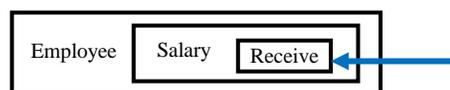

Fig. 3. The Static Model Presented in Terms of a Single Decomposition.

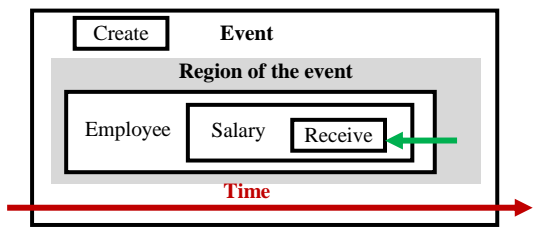

Fig. 4. The Dynamic Model with a Single Event.

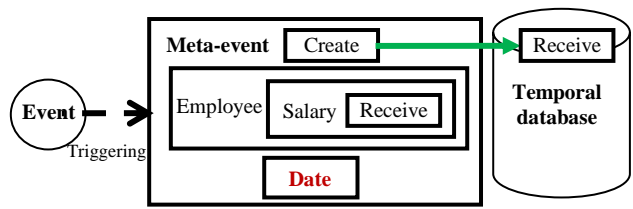

Fig. 5. The Temporal Model Proposed in this Paper.

## II. TM Modeling

The TM model articulates the ontology of the world in terms of an entity that is simultaneously a thing and a machine, called a thimac [13-22]. A thimac is like a double-sided coin. One side of the coin exhibits the characterizations assumed by the thimac; on the other side, operational processes emerge, which provide a dynamism that goes beyond structures or things to embrace other things in the thimac. A thing is subjected to doing (e.g., a tree is a thing that is planted, cut, etc.), and a machine does (e.g., a tree is a machine that absorbs carbon dioxide and uses sunlight to make oxygen). The tree thing and the tree machine are two faces of the tree thimac. A thing is viewed based on Heidegger's [23] notion of thinging. According to Bryant [24], "A tree is a thing through which sunlight, water, carbon dioxide, minerals in the soil, etc., flow. Through a series of operations, the machine transforms those flows of matter, those other machines that pass through it, into various sorts of cells." A thing is a machine, and a machine is a thing. A machine facilitates the movement of things; simultaneously, as a machine, it is a thing in its processual mode.

The simplest type of machine is shown in Fig. 6. The actions in the machine (also called stages) are as follows:

**Arrive:** A thing moves to another machine.

**Accept:** A thing enters a machine. For simplification, we assume that all arriving things are accepted; hence, we can combine the arrival and acceptance of the thing into the receive stage.

**Release:** A thing is marked as ready to for transfer outside the machine (e.g., in an airport, passengers wait to board after passport clearance).

**Process:** A thing is changed in form, but no new thing results.

**Create:** A new thing is born in a machine.

**Transfer:** A thing is input into or output from a machine.

Additionally, the TM model includes storage and triggering (denoted by a dashed arrow in this study's figures), which initiates a flow from one machine to another. Multiple machines can interact with each other through the movement of things or triggering stages. Triggering is a transformation from one series of movements to another (e.g., electricity triggers cold air).

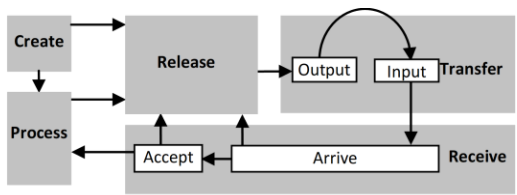

Fig. 6. The Thinging Machine.

### III. TM MODELING EXAMPLE

Because the subject of temporality in this paper is a fourth-level notion in the TM model, this section builds a solid basis for understanding the model before treating the topic of temporal data.

Etzion and Niblett [25] presented a fast flower-delivery system specification as a case study of an event-processing scheme. The case study involves a consortium of flower stores that have an agreement with local independent van drivers to deliver flowers from the stores to their destinations.

When a store gets a flower delivery order, it creates a request, which is broadcast to relevant drivers within a certain distance from the store, with the time for pickup (typically now) and the required delivery time. A driver is then assigned, and the customer is notified that a delivery has been scheduled. The driver makes the pickup and delivery, and the person receiving the flowers confirms the delivery time by signing for it on the driver's mobile device. The system maintains a ranking of each individual driver based on his or her ability to deliver flowers on time. Each store has a profile, which can include a constraint on the ranking of its drivers. The profile also indicates whether the store wants the system to assign drivers automatically or whether it wants to receive several applications and then make its own choice. [21].

Fig. 7 shows the corresponding static TM model. The diagram includes the user (circle 1), the store (2), the driver (3), and the person who receives the flowers (4). The other parts of the diagram model the system. The following occurs accordingly:

- The user creates the order (4), which flows to the store (5), where a minimum ranking requirement (6) is added to the user order to form a delivery order (7).

- The delivery order flows to the system (8) to be processed (9) and is sent to a submachine (10), which extracts a subset of drivers who satisfy the minimum ranking requirement. In this machine, the delivery request is processed (11) to extract the minimum ranking (12), which is compared (13) with driver ranks (14) coming one at a time (15) from the file of all ranked drivers (16). Through this comparison, a file of qualified (minimum ranking) drivers is constructed (17).

- Another machine (18) identifies drivers who are currently in the region by processing.

    - the list of qualified drivers (17);

    - the delivery location, extracted from the delivery request (19); and.

    - the current location of the qualified driver (20). The current locations of drivers (21) are updated continuously via satellite (22).

- If a driver is located in a nearby region, then a bid request is constructed (23) and sent to that driver (24).

- Drivers who receive bid requests respond by creating a delivery bid (25), which flows to the system (26) to be collected with other delivery bids (27).

- When bid requests are sent, a timing machine (28) is constructed to set a deadline of 2 minutes (29). At the end of the 2 minutes, the accumulated delivery bids (27) are processed (30). If there are no bids, then an alert is generated (31) and sent to the store (32) and the system manager (33).

- If there are bids and the store policy is to select the assigned driver manually, then a list of top bidders is generated (34) and sent to the store (35), and a deadline for the response is set (36). If there is no response by the deadline, then, as before, an alert is sent to the store and the system manager (37).

- An assignment of a driver is either created by the system (38) or received from the store (39) and sent to the driver (40).

- The driver goes to the store (41) and picks up (42) the flowers (43). A confirmation of this (44) is sent to the system.

- The driver with the flowers (45) drives to the person who ordered the flowers (46), and a confirmation of the delivery is sent to the system (47). Details such as about how the confirmation is sent through signing the driver's mobile device are not included, due to space considerations.

Additionally, the rest of the case study, which involves updating the drivers' ranks, is not modeled, to limit the model to one page. Some other simplifications were also applied, such as lumping all alerts and all confirmations together.

Fig. 8 shows the events in the model, which were developed as a layer over the static model in Fig. 7.

Event 1 ($E_1$): The user submits an order for flowers to the store.

Event 2 ($E_2$): The store constructs a delivery request that includes minimum driver rankings.

Event 3 ($E_3$): The delivery request flows to the system, in which a list of qualified drivers is produced.

Event 4 ($E_4$): A qualified driver in the nearby region is identified.

Event 5 ($E_5$): A bid request is generated and sent to the qualified driver in the nearby region.

Event 6 ($E_6$): A time deadline (2 minutes) is initiated to receive delivery bids.

Event 7 ($E_7$): The driver formulates a delivery bid, which flows to the system to be stored with all other delivery bids.

Event 8 ($E_8$): The time deadline (2 minutes) to receive delivery bids expires.

Event 9 ($E_9$): The list of all delivery bids is processed.

Event 10 ($E_{10}$: The bid receives no bidders.

Event 11 ($E_{11}$): An alert is generated and sent to the store and system manager.

Event 12 ($E_{12}$): There are bidders.

Event 13 ($E_{13}$): The top delivery bids are selected.

Event 14 ($E_{14}$): The top delivery bids are sent to the store.

Event 15 ($E_{15}$): The timing is set for the store to select a driver.

Event 16 ($E_{16}$): The store selection of a driver passes the deadline.

Event 17 ($E_{17}$): The system selects a driver.

Event 18 ($E_{18}$): The store selects a driver.

Event 19 ($E_{19}$): The selection is sent to the driver.

Event 20 ($E_{20}$): The driver goes to the store and picks up the flowers

Event 21 ($E_{21}$): A confirmation is sent to the system.

Event 22 ($E_{22}$): The driver delivers the flowers to the customer.

Fig. 9 shows the behavioral model of this example. Note that the granularity of events depends on the modeler. For example, Event 3 can be refined further into six events:

Event 3a ($E_{3a}$): The delivery request is inputted to the submachine.

Event 3b ($E_{3b}$): The list (file) of ranked drivers is processed.

Event 3c ($E_{3c}$): A single record of a ranked drivers is retrieved from the file.

Event 3d ($E_{3d}$): The retrieved record flows for comparison with the requested minimum ranking.

Event 3e ($E_{3e}$): If the record does not satisfy the minimum requirement, then it is ignored, and the next record is retrieved from the file containing the ranked drivers.

Event 3f ($E_{3f}$): If the record satisfies the minimum requirement, then it is added to the file of qualified drivers.

The detailed chronology of events for Event 3 can be developed in the same way as in the main model.

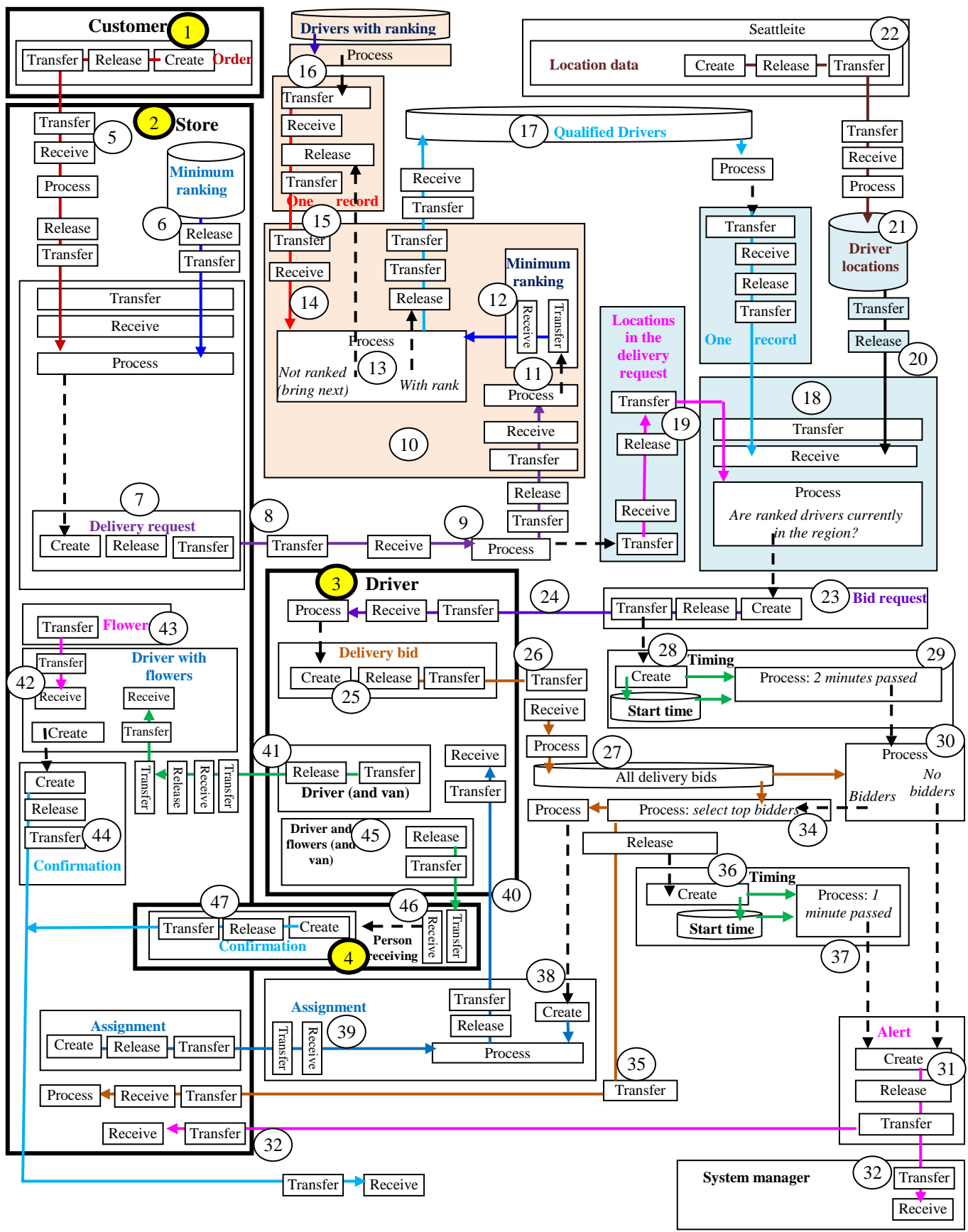

Fig. 7. The Static TM Model of the Consortium of Flower Stores.

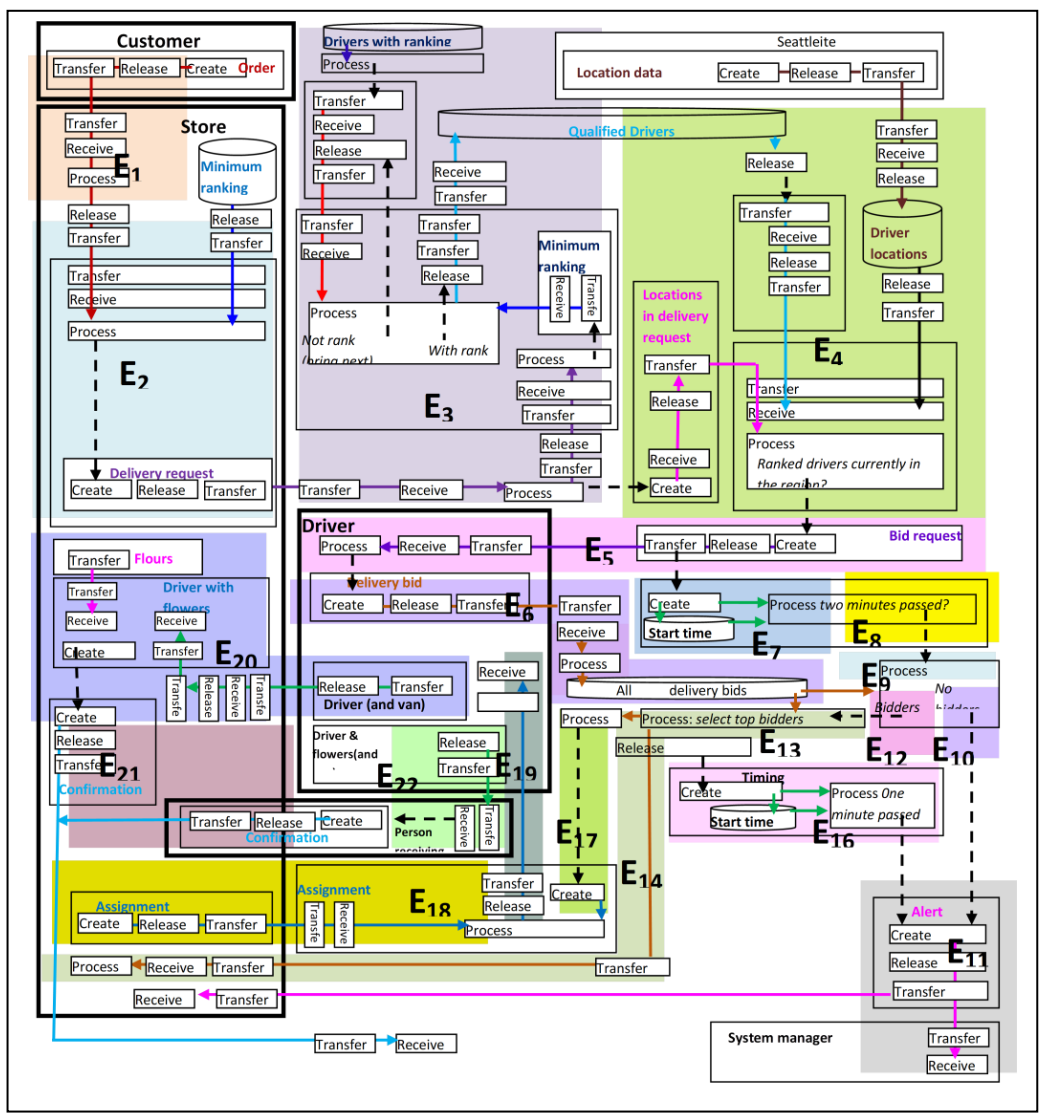

Fig. 8. The Dynamic TM Model of the Flower Store Consortium.

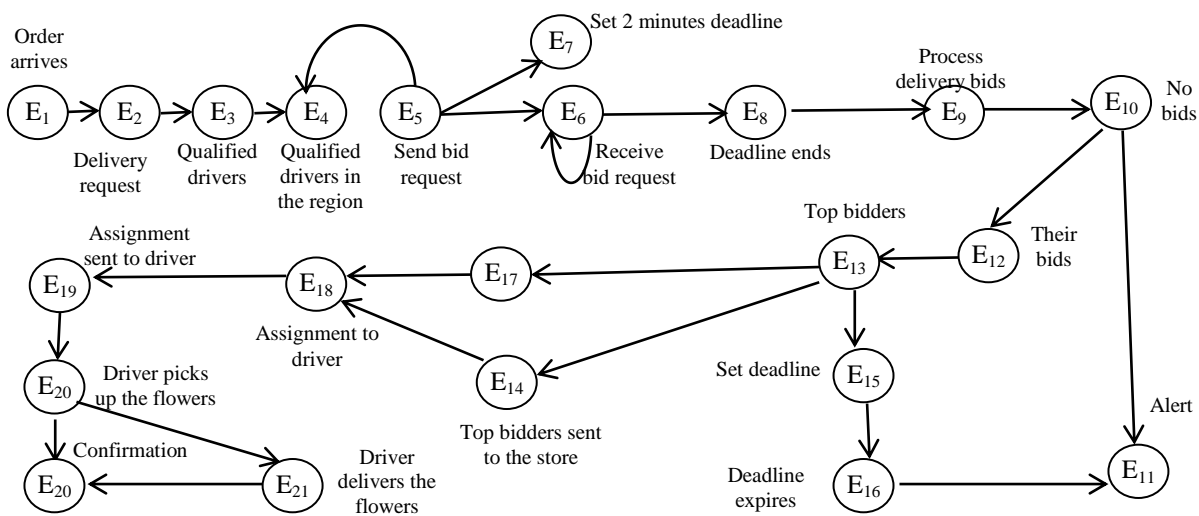

Fig. 9. The Behavioral Model of the Flower Stores.

## IV. Modeling a Bank

According to El Hayat et al. [12], the emergence of temporal databases calls for new, efficient visual-modeling techniques to facilitate the design of temporal objects. El Hayat et al. [12] used extended UML and the OCL to produce an elaborate conceptual schema that allows for defining the restrictions and constraints that contain the duplicate and complex expressions in a temporal database. The proposed temporal UML/OCL model is based on bitemporal data, which translate into a temporal object-relational database for tracking historical information efficiently. El Hayat et al. [12] provided the class diagram of a banking system that includes bitemporal data to make records of the history of the data and transaction operations.

### A. Static Model

This paper lacks the space to describe El Hayat et al.'s [12] model. We will take their example, with some minor simplifications, as an example of a TM conceptualization of a temporal database. Fig. 10 shows the static TM model of the involved banking system. The simplification of El Hayat et al.'s [12] model involves such changes as eliminating some attributes of the customer's address, such as their city and postal code, while keeping their ID, name, and address. The purpose is to remove redundant types of attributes without changing the kinds of data in the model. Additionally, we will focus on four regions in the model that involve the temporality of data: transfer, withdrawal, deposit, and loan transactions.

In Fig. 10, before a customer (circle 1) requests a service, he or she provides an account number (2), which flows to the system, (3) where it is validated (4) with the account number stored in the system (5 and 6).

*a) Loan service:* A customer requests a loan (7); hence, the request moves to the system (8 and 9), where it is processed (10). Assuming the loan is approved, the amount is extracted (11) from the request, along with a generated loan number (12) that flows to the loan subsystem (13), where they and the account number (14) are processed (15) to trigger the creation (16) of a loan record, which is stored in the loan database (17).

The temporal data of such activities will be handled in the events model.

*b) Transactions (18):* A customer requests a transaction (19); hence, the request moves to the system (20 and 21), where it is processed (22). Additionally, the customer provides the transaction request (22), which flows to the system (23). Based on the transaction type (21), the input amount is directed (24) to the transfer (25 and 26), withdraw (27 and 28), or deposit module (29 and 30). The account (31) is updated according to the type of transaction.

*c) Transfer:* The account is retrieved (32) to be received in the transfer module (33) and is processed along with the input amount (34) to create the new account value (35), which flows as the new account value (36).

*d) Withdraw:* The account is retrieved (37) to be received in the transfer module (38) and is processed along with the input amount (39) to create the new account value (40), which flows as the new account value (41).

*e) Deposit:* The account is retrieved (42) to be received in the transfer module (43) and is processed along with the input amount (44) to create the new value of the account (45) that flows as the new value of the account (46).

In this scenario, we ignore the modeling of processes such as checking whether sufficient funds exist for withdrawal or the mechanism of transferring from the account to the intended destination. These processes can be added easily, but the purpose here is not to demonstrate the TM modeling, which was the purpose of the previous section's example. Rather, the purpose of the current example is to show the temporal features of TM modeling by registering the times of account values for multiple transactions. This will be the function of the TM events model.

### B. Dynamic Model

Fig. 11 shows the events model of the bank operations, in which we identify the following events.

Event 1 ($E_1$): The customer requests a loan.

Event 2 ($E_2$): The request flows to the system, where it is approved and sent to the loan module.

Event 3 ($E_3$): The loan amount, number, and account number are processed.

Event 4 ($E_4$): A record of the loan amount, number, and account number is created and stored.

Event 5 ($E_5$): The customer requests a transaction service and gives the amount related to the transaction.

Event 6 ($E_6$): The system determines that the requested service is a transfer and sends the transaction to the transfer module.

Event 7 ($E_7$): The current balance value is retrieved and is processed along with the transfer amount in the transfer module.

Event 8 ($E_8$): The new balance value is created in the transfer module.

Event 9 ($E_9$): The balance value is updated.

Event 10 ($E10_4$): The system determines that the requested service is a withdrawal and sends the transaction to the withdraw module.

Event 11 ($E_{11}$): The current balance value is retrieved and, along with the withdrawal amount, is processed in the withdraw module.

Event 12 ($E_{12}$): The new balance value is created in the withdraw module.

Event 13 ($E_{13}$): The system determines that the requested service is a deposit and sends the transaction to the deposit module.

Event 14 ($E_{14}$): The new balance value is created in the deposit module.

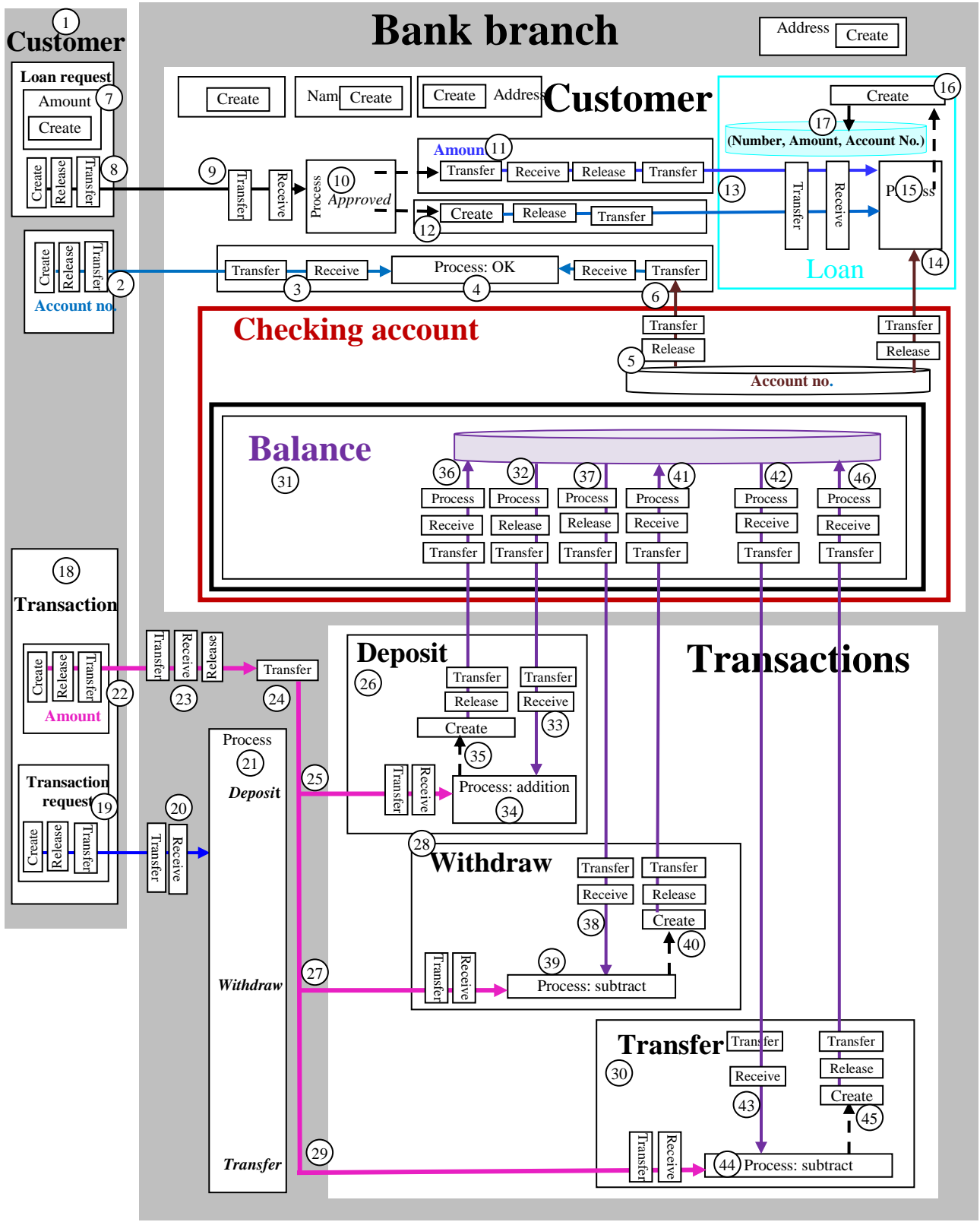

Fig. 10. The Static TM Model of the Bank.

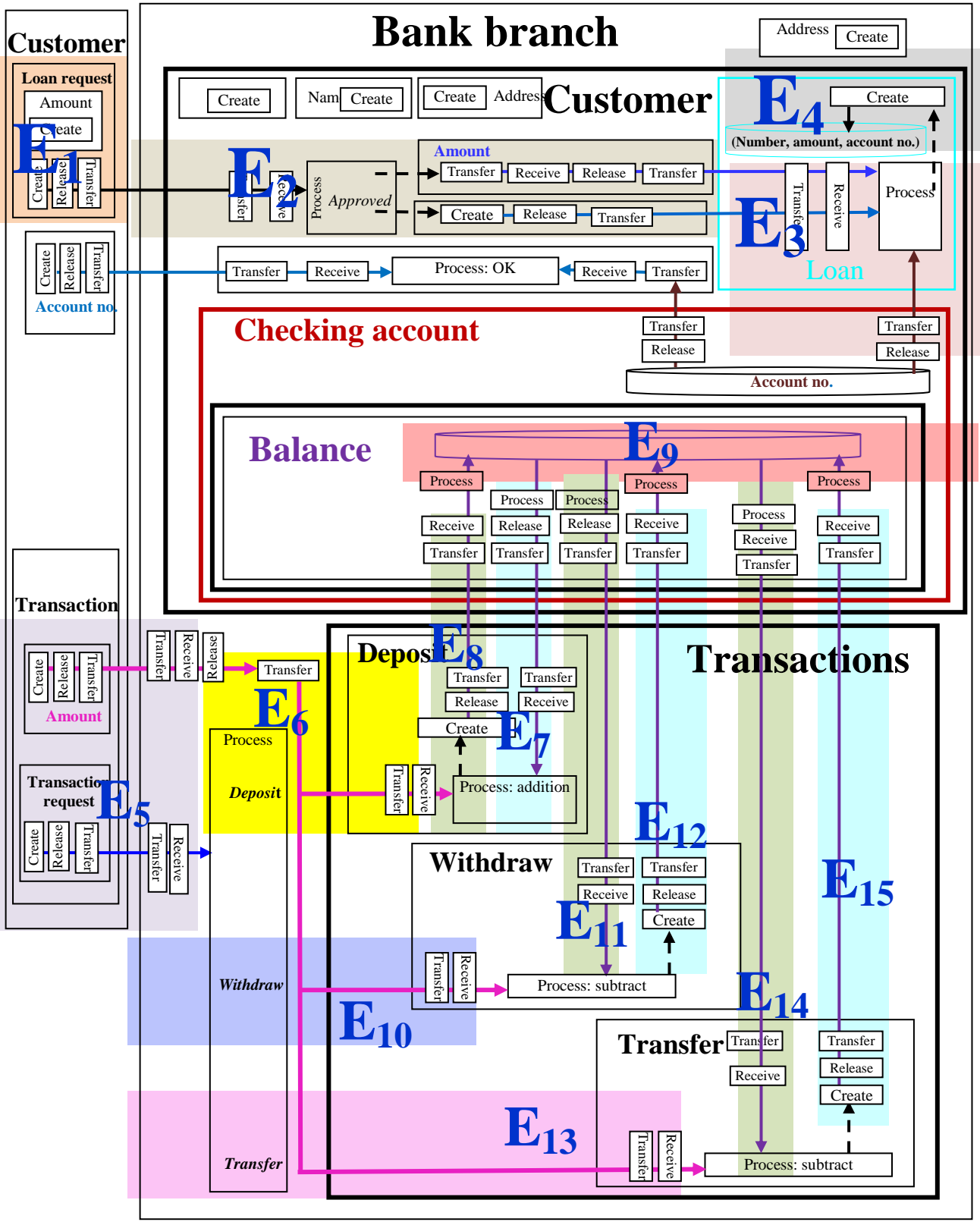

Fig. 11. The Dynamic TM Model for Bank-System Management.

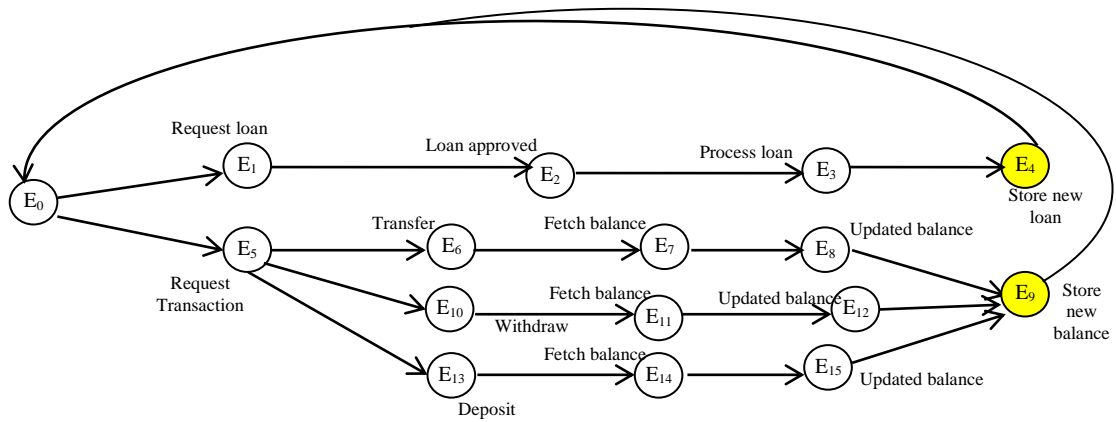

Fig. 12. The TM behavioral of the Bank Services.

Fig. 12 shows the behavioral model of the bank operations.

*C. Temporal Model*

We will focus on the values of the loans and balances at different times. Fig. 13 shows a file being built for the temporal balance value. This change in the balance value occurs at event $E_9$. Accordingly, when event $E_9$ occurs, it triggers a meta-event (an event that is generated by an event), denoted as $ME_9$. $ME_9$ creates a record of $E_9$ that contains data about the time of $E_9$, the new balance value, and the account number. In general, the event time may include different types of time-related data (e.g., the start time, end time, duration, and urgency). Thus, a temporal file of changes to the customer's balance is updated each time a new value is calculated and stored. A similar file can be constructed for loans.

Such a meta-event notion can be applied not only to collect a history of certain data but also to all events in the behavioral model in order to form a temporal-monitoring system of historic records of all activities. For example, a history record can be generated of the events and data changes for each customer's loan. Fig. 14 shows an expansion of the system to monitor the total activity involved in the loan service. For example, in the loan service, not only are the loan value and loan's time recorded, but also all related activities (e.g., time the customer submitted the request or processing time) are recorded and stored when registering the new loan. We can generalize this model to a complete monitoring system that records all activities and data changes over time, as shown in Fig. 15.

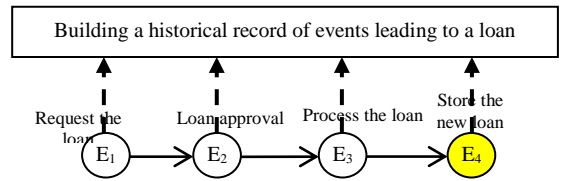

Fig. 14. Historical Record of all Events for a Certain Loan.

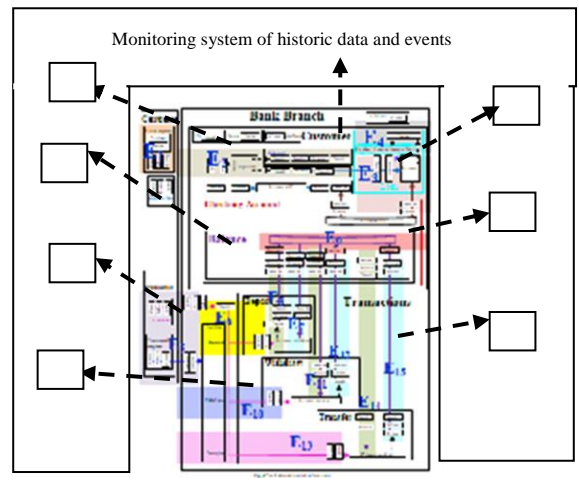

Fig. 15. General Monitoring System for Temporal Data and Changes.

## V. CONCLUSION

In this paper, we have examined a model for temporal databases using an example banking-system-management model that extends UML and OCL. The bank system is re-modeled using TM modeling. The resulting diagrammatic TM specification delivers a new approach to temporality that can be extended to a holistic monitoring system for historic data and events. In this case, a temporal database can be viewed as a restricted monitoring system. One limitation of TM modeling is the complexity of its diagram. However, this apparent complexity originates from the level of granularity of the description. For example, the actions *release*, *transfer*, and *received* can be eliminated under the assumption that the arrow direction will be sufficient to indicate the direction of flows. Future research can be conducted to develop a general monitoring scheme for an entire organization.

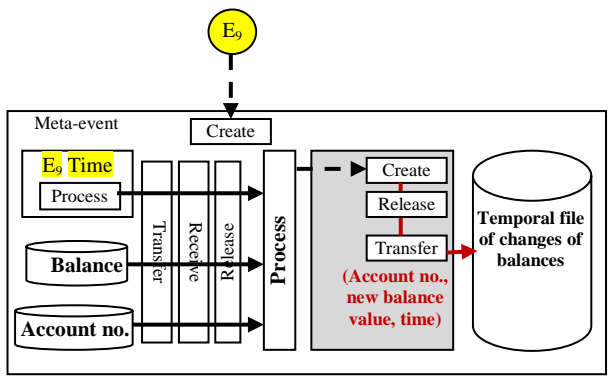

Fig. 13. Generating Temporal Data for Changes in the Balance Value.